\documentclass[conference]{IEEEtran}
\IEEEoverridecommandlockouts
\usepackage{cite}
\usepackage{amsmath,amssymb,amsfonts}
\usepackage{algorithmic}
\usepackage{graphicx}
\usepackage{textcomp}
\usepackage{xcolor, soul}
\sethlcolor{yellow}
\usepackage{comment}

\usepackage{siunitx}
\usepackage{hyperref}
\usepackage[nolist]{acronym}
\usepackage[flushleft]{threeparttable}
\usepackage{placeins}
\usepackage{svg}                                   
\usepackage[percent]{overpic}

\usepackage{colortbl}
\definecolor{Gray}{gray}{0.85}

\usepackage{mathtools}
\usepackage{multirow}
\usepackage{booktabs}

\def\BibTeX{{\rm B\kern-.05em{\sc i\kern-.025em b}\kern-.08em
    T\kern-.1667em\lower.7ex\hbox{E}\kern-.125emX}}
\begin{document}
\bstctlcite{IEEEexample:BSTcontrol}

\newcommand{\RNum}[1]{\uppercase\expandafter{\romannumeral #1\relax}}  
\newcommand{\bfqty}[2]{\text{\bfseries\SI{#1}{#2}}}

\DeclareRobustCommand{\IEEEauthorrefmark}[1]{\smash{\textsuperscript{\footnotesize #1}}}

\DeclareSIUnit\dbi{dBi}                                     
\DeclareSIUnit\dbm{dBm}                                     
\DeclareSIUnit\msInference{ms/inference}              
\title{Skilog: A Smart Sensor System for Performance Analysis and Biofeedback in Ski Jumping}

\author{
\IEEEauthorblockN{
Lukas Schulthess\IEEEauthorrefmark{1}, Thorir Mar Ingolfsson\IEEEauthorrefmark{3}, Marc Nölke\IEEEauthorrefmark{2}, Michele Magno\IEEEauthorrefmark{1}, Luca Benini\IEEEauthorrefmark{3}\textsuperscript{,}\IEEEauthorrefmark{4}, Christoph Leitner\IEEEauthorrefmark{3}
}\\
\IEEEauthorblockA{\IEEEauthorrefmark{1} Center for Project Based Learning, ETH Z{\"u}rich, Z{\"u}rich, Switzerland}
\IEEEauthorblockA{\IEEEauthorrefmark{2} PZN, Krakow, Poland}
\IEEEauthorblockA{\IEEEauthorrefmark{3} Integrated Systems Laboratory, ETH Z{\"u}rich, Z{\"u}rich, Switzerland}
\IEEEauthorblockA{\IEEEauthorrefmark{4} DEI, University of Bologna, Bologna, Italy}
\vspace{-1cm}
\thanks{Corresponding author: \{christoph.leitner\}@iis.ee.ethz.ch}
}

\maketitle



\begin{abstract}

In ski jumping, low repetition rates of jumps limit the effectiveness of training. Thus, increasing learning rate within every single jump is key to success. A critical element of athlete training is motor learning, which has been shown to be accelerated by feedback methods. In particular, a fine-grained control of the center of gravity in the in-run is essential. This is because the actual takeoff occurs within a blink of an eye ($\sim$  $\mathbf{300\mskip3mu}$ms), thus any unbalanced body posture during the in-run will affect flight.

This paper presents a smart, compact, and energy-efficient wireless sensor system for real-time performance analysis and biofeedback during ski jumping. The system operates by gauging foot pressures at three distinct points on the insoles of the ski boot at $\mathbf{100\mskip3mu}$Hz. Foot pressure data can either be directly sent to coaches to improve their feedback, or fed into a \ac{ML} model to give athletes instantaneous in-action feedback using a vibration motor in the ski boot.
In the biofeedback scenario, foot pressures act as input variables for an optimized XGBoost model. We achieve a high predictive accuracy of $\mathbf{92.7\mskip3mu}$\% for center of mass predictions (dorsal shift, neutral stand, ventral shift).
Subsequently, we parallelized and fine-tuned our XGBoost model for a RISC-V based low power parallel processor (GAP9), based on the \ac{PULP} architecture. We demonstrate real-time detection and feedback ($\mathbf{0.0109\mskip3mu}$ms/inference) using our on-chip deployment.
The proposed smart system is unobtrusive with a slim form factor ($\mathbf{13\mskip3mu}$mm baseboard, $\mathbf{3.2\mskip3mu}$mm antenna) and a lightweight build ($\mathbf{26\mskip3mu}$g). Power consumption analysis reveals that the system's energy-efficient design enables sustained operation over multiple days (up to 300 hours) without requiring recharge. 

\end{abstract}

\vspace{1mm}

\vspace{-1mm}

\section{Introduction}\label{sec:introduction}

\makeatletter
\def\ps@IEEEtitlepagestyle{
  \def\@oddfoot{\mycopyrightnotice}
  \def\@evenfoot{}
}
\def\mycopyrightnotice{
  \begin{minipage}{\textwidth}
  \centering \scriptsize
  Copyright~\copyright~2023 IEEE. Personal use of this material is permitted. Permission from IEEE must be obtained for all other uses, in any current or future media, including\\reprinting/republishing this material for advertising or promotional purposes, creating new collective works, for resale or redistribution to servers or lists, or reuse of any copyrighted component of this work in other works.
  \end{minipage}
}
\makeatother

Professional sports are fiercely competitive. In ski jumping, for example, even small improvements in the take-off phase can make a decisive difference between victory and defeat \cite{elfmark_2019}. Within the short time of a jump (less than 10 seconds \cite{muller_2006}), athletes must learn to solve complex motor control and optimization problems while being exposed to harsh environmental conditions, e.g., wind, snow, and low temperatures \cite{bessone_2021, muller_determinants_2009}. Hence, requiring a high level of physical and mental fitness from athletes \cite{raschner_2013}. 

In ski jumping, the actual act of jumping (the leap) must be mastered and optimized in a very short period of time (seconds). Athletes approach take-off tables at run-up speeds of up to \SI{25}{\meter/\second} and perform the entire take-off motion within approximately \SI{300}{\milli\second} \cite{j:schwameder2008}. Consequently, any unbalanced body posture during the in-run (preparation for take-off) and the take-off itself can cause significant technical challenges and lead to poor performance of the jump. 
Therefore, athletes with the ability to fine-tune their center of gravity during the in-run phase can improve jumping performance \cite{j:janura2010}. For this reason, coaches and federations are exploring new technology-based solutions to gain a better awareness of the body's center of gravity to accelerate the motor learning of the correct jumping technique \cite{j:pustisek_2021}.

State-of-the-art performance assessment in ski jumping is usually accomplished with the help of video footage \cite{j:murakami2014}. Jumps are recorded at the coaching tower and evaluated by trainers, who provide verbal feedback to improve the athletes' posture and dynamics in the next jump. 
Studies have shown that augmented feedback can improve motor learning over traditional verbal or visual approaches \cite{ j:mulder1985, j:sigrist2015}. In ski jumping, for example, biofeedback methods have so far been used for stress management during pre-start \cite{b:pointner2014}, but, to the best of our knowledge, there are no wearable devices for real-time monitoring and biofeedback during the actual ski jumping workout. 
Another limiting factor in motor learning of ski jumping is the limited number of possible repetitions \cite{j:link_2021}. In contrast to cyclic sports, the ratio between the jump duration and the time to prepare for the next attempt is large and usually ranges up to several minutes. This significantly reduces the number of repetitions and consequently, the amount of time able to be spent practicing ski jumping over the course of a ski jumping career \cite{j:link_2021}. Therefore, an assistive device providing feedback on the quality of motion could accelerate motor learning.
\begin{table}[]
\renewcommand{\arraystretch}{1}
  \centering
  \caption{The largest ski jumping facilities}\label{tab:jumping_facilites}
  \vspace{-.2cm}
  {
    \footnotesize
    \begin{tabular}{@{}lll@{}}
        \toprule
            Track & Hill size & Location \\
        \midrule
            Vikersundbakken & \SI{240}{\meter} & Vikersund, Norway \\
            Letalnica bratov Gorišek & \SI{240}{\meter} & Planica, Slovenia \\
            Kulm-Skiflugschanze & \SI{235}{\meter} & Bad Mitterndorf, Austria \\
            Heini-Klopfer-Schanze & \SI{235}{\meter} & Oberstdorf, Germany \\
            Čerťák & \SI{210}{\meter} & Harrachov, Czech Republic \\
            Copper Peak & \SI{160}{\meter} & Ironwood, USA \\
      \bottomrule
    \end{tabular}
  }
  \vspace{-0.6cm}
\end{table}

\begin{figure}[b]
    \centering
    \vspace{-7mm}
    \begin{overpic}
        [width=1\columnwidth]{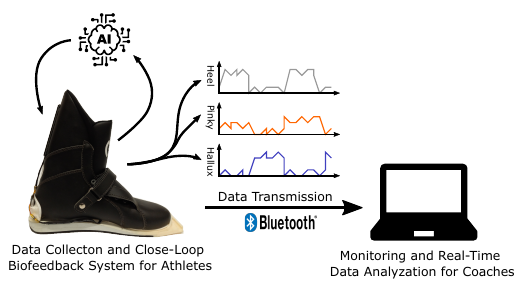}
        \put(27,27){(1)}         
        \put(10,49){(2)}        
    \end{overpic}
    \vspace{-7mm}
    \caption{The proposed training system supports two modes: (1) streaming raw data in real-time for post-action in-depth performance analysis by the coach, and (2) performing on-the-edge training analysis and providing instantaneous feedback to the athlete using a vibration motor.}
    \label{fig:system_overview}
\end{figure}
Current on-body sensing technologies in ski jumping are mainly used to quantify flight trajectory parameters with the help of \acp{IMU} \cite{j:chardonnens_2012}. These devices are either strapped onto the athlete's body \cite{j:chardonnens_2012, baechlin_2010}, mounted onto skis \cite{j:logar_2015} or attached to the bindings \cite{j:link_2021}. 
Real wearable sensing, however, imposes strict design requirements which are not met by currently used devices \cite{c:frey2022}. They must be energy efficient (to guarantee long run-time), unobtrusive, and nearly imperceptible (to not interfere with natural movement behaviors and the jumping technique), and in particular, they must be equipped with a wireless link (for real-time data analyses, e.g., on the coaching tower) \cite{j:hribernik_2022, j:rana_2021}. 
Guaranteeing reliable communication in such a scenario is quite challenging. The data link needs to support the largest jumping constructions that exceed distances of \SI{240}{\meter} from the caching tower (usually located in the middle of the jumping facility) up to the end of the slope, \autoref{tab:jumping_facilites} \cite{w:ski_slopes}.

Consumer sports on the other hand, already make substantial use of wearable wireless technology, such as smartwatches and fitness trackers for pervasive activity monitoring \cite{w:fitbit} (e.g. step count, heart rate, VO2max, etc.), biomechanical performance assessment \cite{w:garmin} (eg. balance, ground reaction forces cadence, oscillation) as well as for biofeedback \cite{w:biosign}. 
However, in ski jumping, and specifically in the improvement of the center of gravity position in the in-run track, such generic devices do not provide enough application-specific data to improve motor learning \cite{j:hribernik_2022, j:pustisek_2021}. 

In this context, we present a tiny, smart, and low-power wireless sensing system for real-time performance analysis and biofeedback during ski jumping. Foot pressure distributions are sampled at three measurement points in the insole of a ski boot (\SI{100}{\hertz} sampling frequency) and can be streamed to the coaching tower during jumping. 
Moreover, our system is able to determine the center of gravity by measuring the foot's pressure distribution at three distinctive positions (hallux, pinky, and heel) with an accuracy of \SI{92.7}{\percent} using a lightweight machine learning model (\SI{38.29}{\kilo\byte} memory footprint) deployed on a novel RISC-V based low power parallel processor (GAP9), based on the \ac{PULP} architecture. 
The system provides classification results in real-time (\SI{0.0114}{\msInference}) which can serve as input features for appropriate feedback modalities (e.g. haptic, visual, or bioelectronic) for athletes. In addition, form factor (height profile: baseboard \SI{13}{\milli\meter}, antenna \SI{3.2}{\milli\meter}) and weight of our system (\SI{26}{\gram} incl. battery and antenna) are small and light and thus below the notification threshold of ski jumpers. Exploiting \ac{BLE} coded PHY as \ac{RF} protocol guarantees a reliable data link over the whole training facility \cite{j:janssen_2020}.

\section{Material and Methods}\label{sec:concept}
The proposed system consists of a modified ski boot in which we integrated three piezoresistive \ac{FSR} sensors measuring the pressure distribution on the foot soles of ski jumpers. To evaluate the proposed sensor subsystem we record raw data from pressure sensors using three dedicated \acp{ADC} with a 12-bit resolution on a \ac{SoC}. The recorded data can be sent in real-time over \ac{BLE} coded PHY to the coaching tower during jumping (\autoref{fig:system_overview}, (1)). 
In parallel, an XGBoost classifier directly runs on the system to perform low-latency and energy-efficient analysis of raw pressure data. The \ac{ML} model provides instant predictions of the athlete's center of gravity and biofeedback is provided by a vibration motor (\autoref{fig:system_overview}, (2)). We base the classification task on XGBoost, a parallelized and highly optimized manifestation of the Gradient Boosted Tree algorithm \cite{b:hastie_2009}. The justification for adopting an XGBoost classifier within this paper emerges from its previously demonstrated efficacy in bio-signal classification tasks in a fast and energy-efficient manner \cite{biocas_thorir, c:ingolfsson2022}.

\begin{figure}[]
    \centering
    \vspace{-7mm}
    \begin{overpic}
        [width=1\columnwidth]{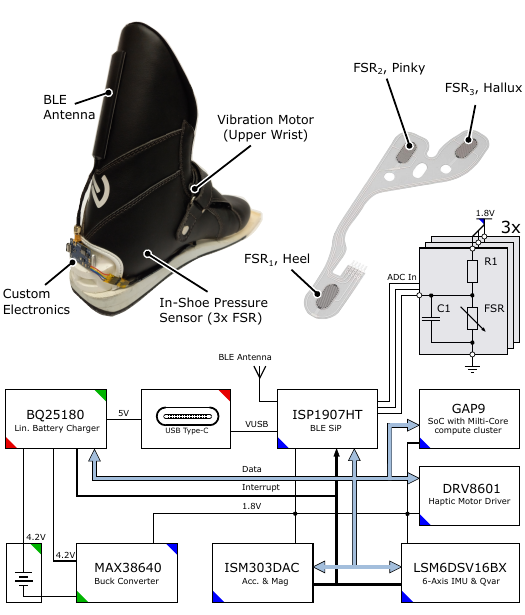}
        \put(0,92){(a)}         
        \put(55,92){(b)}        
        \put(0,39){(c)}         
    \end{overpic}
    \vspace{-7mm}
    \caption{(a) Overview of the modified ski boot, the antenna, the custom electronics mounted just above the heel clamp, and piezoresistive shoe insoles placed inside the boot. (b) Picture of the shoe insole sensor with three \acp {FSR} for measuring pressure at the heel, pinky, and hallux. (c) High-level architecture of the custom electronics with three \ac{FSR} frontends.}
    \label{fig:skilog_high_level}
    \vspace{-4mm}
\end{figure}
\subsection{Hardware Architecture}
The proposed smart sensing system is designed to accommodate multiple sensors, microcontrollers, and a battery, yet with a size and weight that is imperceptible to the athlete. Its light weight of only \SI{26}{\gram}, including battery and the antenna, and the low height profile of \SI{13}{\milli\meter} for the device itself and \SI{3.2}{\milli\meter} for the antenna, minimize impacts on athletes.

\autoref{fig:skilog_high_level} (c) shows a simplified block diagram of the data logger.
The core of the sensor node is the \textit{ISP1907HT} (Insight SiP), a \ac{SiP} based on the nRF52833 \ac{SoC} (Nordic Semiconductor). It integrates \ac{RF} matching, as well as an optional internal antenna and both, a \SI{32}{\kilo\hertz} and \SI{32}{\mega\hertz} crystal, offering a great balance between light weight, size, and cost. Moreover, we have integrated a RISC-V-based microcontroller, tailored for tiny ML applications, into our system design: the \textit{GAP9} (Greenwaves) \cite{w:greenwaves2022} acts as a co-processor to accelerate ML workloads while keeping power dissipation low \cite{w:mlcommons2022}. The processor has ten cores and is based on the RISC-V instruction set architecture. 
The complete system is supplied from a single lithium-polymer battery of type \textit{ICP521630PM} (Renata Batteries) with a total capacity of \SI{240}{\milli\ampere{}\hour}. An integrated step-down converter \textit{MAX38640} (Analog Devices) generates the system operating voltage of \SI{1.8}{\volt}.
To measure the weight distribution on the athlete's foot, a shoe insole sensor of type \textit{RP-INS-3Z} (Taidacent) was integrated into the ski boot (ref. \autoref{fig:skilog_high_level} (a), (b)); three \acp{FSR}, one the heel, the pinky and the hallux.
The external antenna \textit{A2O5RPSMA} (Data Alliance) was mounted on the shaft of the boot. 

Its waterproof housing and low height profile of \SI{3.2}{\milli\meter} meet the design requirements for ski jumping data loggers. Biofeedback is given over a small vibration sensor of type \textit{VZ30C1T8460002L} (Vibronics) which is controlled over the haptic motor driver \textit{DRV8601} (Texas Instruments). 

In addition, our system features a \textit{ISM303DAX} (ST Microelectronics), a high-performance 3D accelerometer and 3D magnetometer, as well as a 6-axis \ac{IMU} of type \textit{LSM6DSV16BX} (ST Microelectronics) to collect data about the ski's orientation and in-flight angle during future in-field data collection.

\subsection{Embedded implementation of the body position classifier}

\begin{figure}[h]
    \centering
    \begin{overpic}
        [width=1\columnwidth]{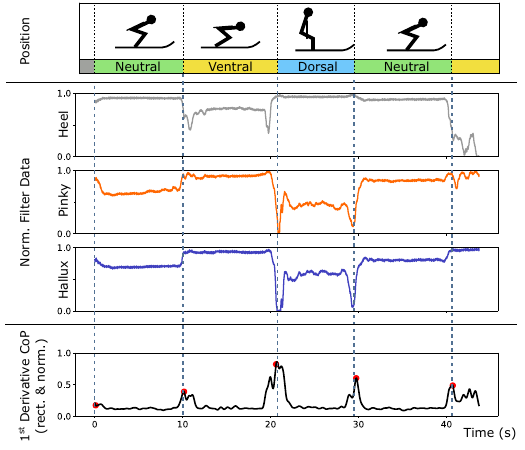}
    \end{overpic}
    \vspace{-6mm}
    \caption{This figure gives an insight into the laboratory recordings of foot pressure data. The top row shows the body position of the ski jumper, which corresponds to our classification label. The graphs in the middle part show filtered foot pressure data on heel, pinky, and hallux. The bottom figure shows the normalized and rectified signals of the first derivative of the center of pressure used to automatically detect the labels.}
    \label{fig:datacollection}
\end{figure}

From raw ADC signals collected at the pressure measurement points in the ski boot (\autoref{fig:datacollection}), we derived the center of pressure of the foot. The neutral position was defined by the centroid of the three contact points (\autoref{fig:skilog_high_level} (b)). We identified dorsal and ventral shifts from the neutral position that resulted from a displacement of the center of gravity. 

For model training, we first collected a foot pressure dataset simulating different body positions of ski jumpers. Subsequently, we adopted an XGBoost classifier to determine the body position in real time. The XGBoost has the added benefit of being a low-latency and energy-efficient model. We deployed our machine learning model on a GAP9 processor, which has been shown to be at least one order of magnitude more energy efficient than similar ARM-based solutions \cite{c:ingolfsson2021}. 

\subsubsection{Dataset collection and Labelling}
We collected a pressure sensor datatset on one volunteer using the proposed sensor interfaces and a USB 6216 data acquisition system (National Instruments) operated at a sampling frequency of \SI{400}{\kilo\hertz}. We recorded 3 sessions of 80 seconds, during which the subject was asked to either stay neutral or to shift the weight in ventral or dorsal directions every 10 seconds. The recorded raw data were down-sampled to a rate of \SI{100}{\hertz} which corresponds with our ADC setup on the nRF microcontroller. The complete dataset contained 27,066 time samples and each time sample contained 1 data point of every ADC.
%

\subsubsection{Automatic Labeling and Feature Extraction}
For automatic data labeling, we first filtered the raw ADC data using a second-order Butterworth low-pass filter with a cutoff frequency of \SI{50}{\hertz}. We then calculated the time derivative of the centroid (centroid was computed across the three pressure points at the hallux, pinky, and heel, \autoref{fig:skilog_high_level} (b)). Subsequently, we rectified the derived centroid time series and extracted the labels (neutral position, dorsal or ventral displacement) that denote signals between the transition points in the time series (\autoref{fig:datacollection}, bottom diagram).

\subsubsection{Synthesis of the train-test datasets}
For the training and testing of our XGBoost model, we first grouped all measurements by their label. Then we synthesized the dataset for offline training and testing by first splitting the dataset into discrete blocks of 50 samples per ADC\footnote{We used the input length of $50$ samples per pressure sensor as this corresponds to the length of our $16$ bit data buffers on the NRF microcontroller.}.
In the next step, we concatenated the data buffers of each ADC to one temporal super-sample consisting of $150$ input features. This resulted in $167$ super-samples of the "dorsal" class, $194$ of the "neutral position" class, and $178$ super-samples of the "ventral" class.
From this dataset, we randomly selected \SI{80}{\percent} of the data into the training and \SI{20}{\percent} into the test set. We trained our XGBoost model using a logarithmic loss function and set the number of estimators to $270$ ($90$ for each class).

\subsubsection{Edge-deployment}
To deploy the tree ensemble of our XGBoost model onto the GAP9 processor, we utilize a similar methodology as described in \cite{c:ingolfsson2022}. Therefore, an in-house compiler written in Python transfers the XGBoost model's main characteristics (feature array, threshold array, index array) in an automated way to a C lookup table. We then split the computation of the boosters evenly between the cores of the GAP9's computational cluster, and the number of estimators is adapted to be divisible by $9$, such that the computational load is evenly distributed between nine cluster cores. 

\subsection{Power measurements}
For the power analysis of the system, we used the Power Profiler Kit 2 (Nordic Semiconductor) 
to source and measure its consumption at \SI{1.8}{\volt}. Separate measurements for data acquisition and \ac{BLE} transmission have been conducted. For analyzing the inference task, we ran the GAP9 at 240 MHz and measured its power consumption by sourcing the microcontroller and measuring its consumption also at \SI{1.8}{\volt}.

\section{Results and Discussion}\label{sec:results}

\subsection{System Operation}

We determined our systems' total power consumption at \SI{2.52}{\milli\watt}.
Out of this, \SI{848}{\micro\watt} is needed for data acquisition from the FSRs, and 
\SI{1.67}{\milli\watt} is used for raw data streaming over \ac{BLE} at \SI{0}{\dbm}. 
With the selected battery of \SI{240}{\milli\ampere{}\hour} at \SI{3.7}{\volt}, a total lifetime of more than \SI{300}{\hour} can be achieved. This suggests being more than enough to sustain multi-day hill training without recharging.

%
\begin{figure}[]
  \centering
  \includegraphics[width=0.4\textwidth, clip, trim={0 0 0 0}]{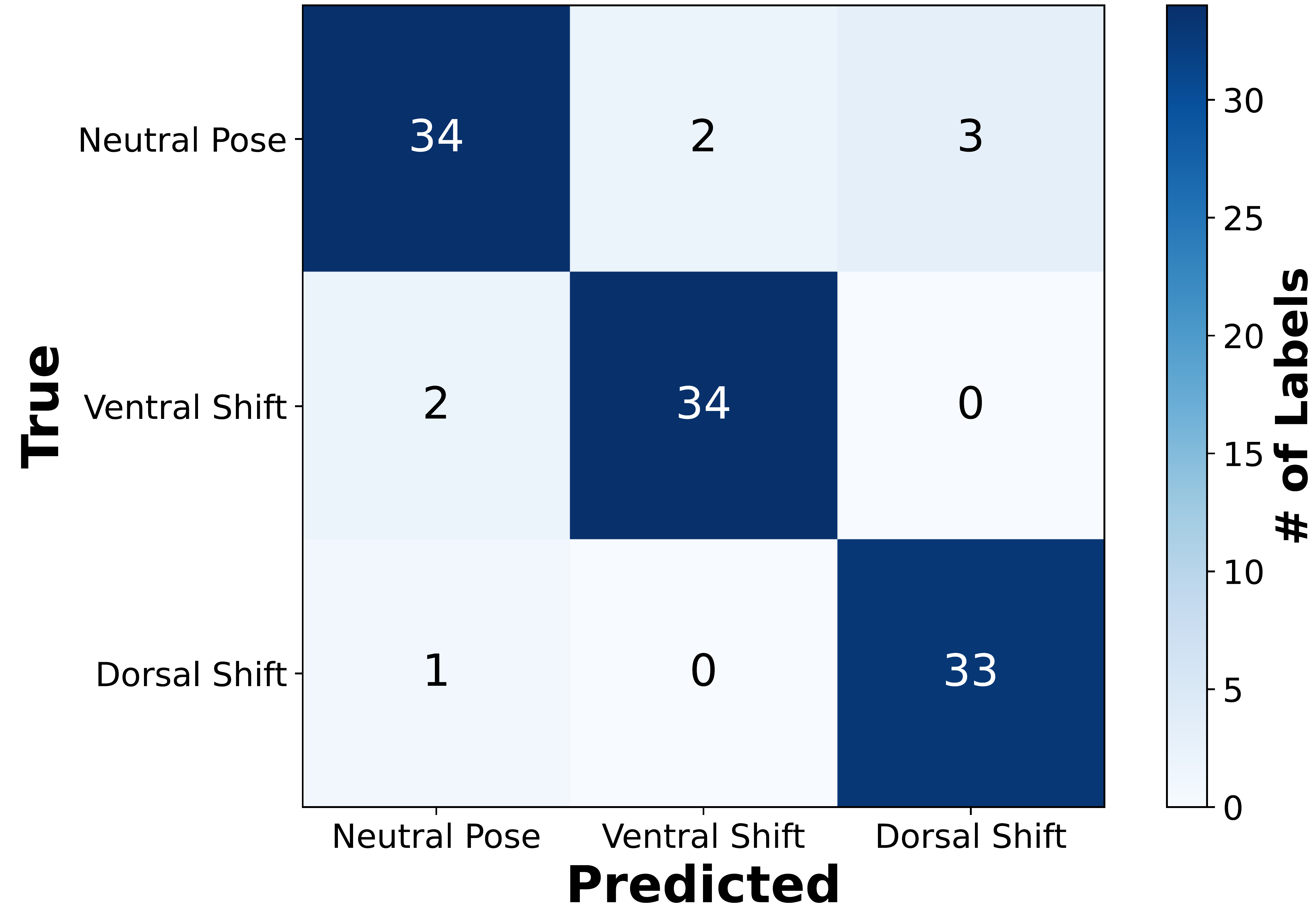}
  \vspace{-1mm}
  \caption{The confusion matrix shows the results of the deployed XGBoost model on GAP9. The test set was fully excluded from network training and consisted of 109 "supersamples" (representing 20\% of the total data set)}
  \vspace{-4mm}
  \label{fig:confusion}
\end{figure}

\subsection{Classification Results on the Edge}
The confusion matrix given in \autoref{fig:confusion} summarizes the classification results for testing our XGBoost model on GAP9. Our model achieves \SI{92.7}{\percent} 
 accuracy in identifying body positions from laboratory-recorded data. The results show that the prediction accuracy on the test set is similar between classes. However, actual center of gravity shifts in experienced ski jumpers are subtle and hardly perceptible to the untrained observer. A transfer to real ski jumping data might therefore require more fine-grained label definitions. Performance numbers of the embedded implementation of our XGBoost model on GAP9 are provided in \autoref{tab:results:summary:gap9}. Low energy numbers (\SI{0.251}{\micro\joule} per inference) and fast inference (\SI{0.0109}{\msInference}) are achieved by distributing the computational load across a nine-core computational cluster. We further explored the potential benefits of such a parallelization approach compared to single-core operation. Our findings highlight a remarkable $6.51\times$ acceleration in inference speed (from \SI{0.071}{\milli\second} to \SI{0.0109}{\milli\second}) and a significant $3.56\times$ energy reduction (from \SI{0.9}{\micro\joule} to \SI{0.25}{\micro\joule}). 
Inherent architectural aspects inevitably limit the speedup factor to reach the theoretical limit of $9\times$. 
These include sharing of four floating-point units by the nine cores, conflicts arising from concurrent access to \ac{TCDM} banks, and instruction misses on the instruction cache.

\begin{table}[t]
\renewcommand{\arraystretch}{1}
  \centering
  \caption{Performance Parameters of XGBoost implementation on GAP9 running at $240$MHz}\label{tab:results:summary:gap9}
  \vspace{-.2cm}
  {
    \footnotesize
    \begin{tabular}{@{}lrrrr@{}}
      \toprule
      Energy cost per inference [$\mu$J] & 0.251  \\
      Time per inference [ms]   & 0.0109\\
      Memory footprint [kB] & 38.29 \\
      \bottomrule
    \end{tabular}
  }
  \vspace{-5mm}
\end{table}

\section{Conclusion}\label{sec:conclusion}
This work presented a novel sensing and training system for body position analyses during ski jumping. Our system can transmit data to coaches and provide direct in-action feedback to athletes by exploiting the advantage of a dual-\ac{SoC} architecture. Our system is the first truly wearable training tool in ski jumping that can provide ski jumpers with a new training experience and could shorten the time for motor learning. 
Moreover, real-time data transmission of biomechanical relevant features can impact TV broadcasting (e.g. by giving viewers more dynamic insights into a ski jump).

The evaluation of the system in a laboratory scenario shows promising results for both, the biofeedback performance and the battery lifetime. The deployment of the XGBoost classification algorithm to GAP9 has successfully 
demonstrated its efficacy in identifying the ski jumpers' body position in an energy-efficient fashion, while simultaneously upholding the stipulations associated with real-time detection and classification.
Future work will involve transferring our hardware and algorithms into the wild by recording a dataset from a large number of subjects. 
We anticipate that model predictions will require further optimization when applied to real-world data.

\FloatBarrier
\begin{acronym}
    \acro{RF}{Radio Frequency}
    \acro{IoT}{Internet of Things}
    \acro{IoUT}{Internet of Underwater Things}
    \acro{UWN}{Underwater Wireless Network}
    \acro{UWSN}{Underwater Wireless Sensor Node}
    \acro{AUV}{Autonomous Underwater Vehicles}
    \acro{UAC}{Underwater Acoustic Channel}
    \acro{FSK}{Frequency Shift Keying}
    \acro{OOK}{On-Off Keying}
    \acro{ASK}{Amplitude Shift Keying}
    \acro{UUID}{Universal Unique Identifier}
    \acro{PZT}{Lead Zirconium Titanate}
    \acro{AC}{Alternating Current}
    \acro{NVC}{Negative Voltage Converter}
    \acro{NVCR}{Negative Voltage Converter Rectifier}
    \acro{FWR}{Full-Wave Rectifier}
    \acro{MCU}{Microcontroller}
    \acro{GPIO}{General Purpose Input/Output}
    \acro{PCB}{Printed Circuit Board}
    \acro{AUV}{Autonomous Underwater Vehicle}
    \acro{IMU}{Intertial Measurement Unit}
    \acro{BLE}{Bluetooth Low Energy}
    \acro{FSR}{Force Sensing Resistor}
    \acro{SiP}{System in Package}
    \acro{SoC}{System on Chip}
    \acro{SpO2}{Oxigen Saturation}
    \acro{PULP}{Parallel Ultra-Low Power}
    \acro{ML}{Machine Learning}
    \acro{ADC}{Analog to Digital Converter}
    \acro{TCDM}{Tightly Coupled Data Memory}
\end{acronym}
\section*{ACKNOWLEDGMENT}
The authors would like to thank Alexander Pointner for his support. Furthermore, this project was supported in part by the Swiss National Science Foundation under the grant agreement 193813 (project PEDESITE) and under grant agreement 209675 (CHIST-ERA project SNOW).

\bibliographystyle{IEEEtranDOI} 
\bibliography{references.bib}

\begin{thebibliography}{10}
\providecommand{\url}[1]{#1}
\csname url@samestyle\endcsname
\providecommand{\newblock}{\relax}
\providecommand{\bibinfo}[2]{#2}
\providecommand{\BIBentrySTDinterwordspacing}{\spaceskip=0pt\relax}
\providecommand{\BIBentryALTinterwordstretchfactor}{4}
\providecommand{\BIBentryALTinterwordspacing}{\spaceskip=\fontdimen2\font plus
\BIBentryALTinterwordstretchfactor\fontdimen3\font minus \fontdimen4\font\relax}
\providecommand{\BIBforeignlanguage}[2]{{%
\expandafter\ifx\csname l@#1\endcsname\relax
\typeout{** WARNING: IEEEtran.bst: No hyphenation pattern has been}%
\typeout{** loaded for the language `#1'. Using the pattern for}%
\typeout{** the default language instead.}%
\else
\language=\csname l@#1\endcsname
\fi
#2}}
\providecommand{\BIBdecl}{\relax}
\BIBdecl

\bibitem{elfmark_2019}
O.~Elfmark \emph{et~al.}, ``Assessment of the steady glide phase in ski jumping,'' \emph{Journal of Biomechanics}, vol. 139, 2022, doi: 10.1016/j.jbiomech.2022.111139.

\bibitem{muller_2006}
W.~Müller, ``The {Physics} of {Ski} {Jumping},'' in \emph{Proceedings of {European} {School} of {High}-{Energy} {Physic}}, Geneva, Switzerland, 2005, doi: http://dx.doi.org/10.5170/CERN-2006-014.269.

\bibitem{bessone_2021}
V.~Bessone and A.~Schwirtz, ``Landing in ski jumping: A review about its biomechanics and the connected injuries,'' \emph{Journal of Science in Sport and Exercise}, vol.~3, pp. 238--248, 07 2021, doi: 10.1007/s42978-020-00096-9.

\bibitem{muller_determinants_2009}
W.~Müller, ``Determinants of {Ski}-{Jump} {Performance} and {Implications} for {Health}, {Safety} and {Fairness},'' \emph{Sports Medicine}, vol.~39, no.~2, pp. 85--106, 2009, doi: 10.2165/00007256-200939020-00001.

\bibitem{raschner_2013}
C.~Raschner \emph{et~al.}, ``Current performance testing trends in junior and elite austrian alpine ski, snowboard and ski cross racers,'' \emph{Sport-Orthopädie - Sport-Traumatologie - Sports Orthopaedics and Traumatology}, vol.~29, no.~3, pp. 193--202, 2013, doi: https://doi.org/10.1016/j.orthtr.2013.07.016.

\bibitem{j:schwameder2008}
H.~Schwameder, ``Biomechanics research in ski jumping, 1991–2006,'' \emph{Sports Biomechanics}, vol.~7, no.~1, pp. 114--136, 2008, doi: 10.1080/14763140701687560.

\bibitem{j:janura2010}
M.~Janura \emph{et~al.}, ``Kinematic characteristics of the ski jump inrun: a 10-year longitudinal study,'' \emph{Journal of Applied Biomechanics}, vol.~26, no.~2, pp. 196--204, 2010, doi: 10.1123/jab.26.2.196.

\bibitem{j:pustisek_2021}
M.~Pustišek \emph{et~al.}, ``The role of technology for accelerated motor learning in sport,'' \emph{Personal and Ubiquitous Computing}, vol.~25, p. 969–978, 12 2021, doi: 10.1007/s00779-019-01274-5.

\bibitem{j:murakami2014}
M.~Murakami \emph{et~al.}, ``High-speed video image analysis of ski jumping flight posture,'' \emph{Sports Engineering}, vol.~17, no.~4, pp. 217--225, 2014, doi: 10.1007/s12283-014-0157-z.

\bibitem{j:mulder1985}
T.~Mulder and W.~Hulstijn, ``Sensory {Feedback} in the {Learning} of a {Novel} {Motor} {Task},'' \emph{Journal of Motor Behavior}, vol.~17, no.~1, pp. 110--128, 1985, doi: 10.1080/00222895.1985.10735340.

\bibitem{j:sigrist2015}
R.~Sigrist \emph{et~al.}, ``Sonification and haptic feedback in addition to visual feedback enhances complex motor task learning,'' \emph{Experimental Brain Research}, vol. 233, no.~3, pp. 909--925, 2015, doi: 10.1007/s00221-014-4167-7.

\bibitem{b:pointner2014}
A.~Pointner and A.~Pointner, \emph{Mut zum Absprung}.\hskip 1em plus 0.5em minus 0.4em\relax Seifert Verlag, 2014, ISBN 3-902924-33-0.

\bibitem{j:link_2021}
J.~Link \emph{et~al.}, ``Experimental validation of real-time ski jumping tracking system based on wearable sensors,'' \emph{Sensors}, vol.~21, no.~23, p. 7780, 2021, doi: 10.3390/s21237780.

\bibitem{j:chardonnens_2012}
J.~Chardonnens \emph{et~al.}, ``Automatic measurement of key ski jumping phases and temporal events with a wearable system,'' \emph{Journal of Sports Sciences}, vol.~30, no.~1, pp. 53--61, 2012, doi: 10.1080/02640414.2011.624538.

\bibitem{baechlin_2010}
M.~Bächlin \emph{et~al.}, ``Ski jump analysis of an olympic champion with wearable acceleration sensors,'' in \emph{International Symposium on Wearable Computers (ISWC) 2010}, Seoul, South Korea, 2010, doi: 10.1109/ISWC.2010.5665851.

\bibitem{j:logar_2015}
G.~Logar and M.~Munih, ``Estimation of joint forces and moments for the in-run and take-off in ski jumping based on measurements with wearable inertial sensors,'' \emph{Sensors}, vol.~15, pp. 11\,258--11\,276, 2015, doi: 10.3390/s150511258.

\bibitem{c:frey2022}
S.~Frey \emph{et~al.}, ``{WULPUS}: a {Wearable} {Ultra} {Low}-{Power} {Ultrasound} probe for multi-day monitoring of carotid artery and muscle activity,'' in \emph{2022 {IEEE} {International} {Ultrasonics} {Symposium} ({IUS})}, Venice, Italy, 2022, doi: 10.1109/IUS54386.2022.9958156.

\bibitem{j:hribernik_2022}
M.~Hribernik \emph{et~al.}, ``Review of real-time biomechanical feedback systems in sport and rehabilitation,'' \emph{Sensors}, vol.~22, no.~8, 2022, doi: 10.3390/s22083006.

\bibitem{j:rana_2021}
M.~Rana and V.~Mittal, ``Wearable sensors for real-time kinematics analysis in sports: A review,'' \emph{IEEE Sensors Journal}, vol.~21, no.~2, pp. 1187--1207, 2021, doi: 10.1109/JSEN.2020.3019016.

\bibitem{w:ski_slopes}
{Ski Jumping Hill Archive}, ``World's largest ski jumpsz,'' \url{https://www.amazon.com/Taidacent-RP-INS-3Z-Analysis-Pressure-Sensitive/dp/B07L3MSKD8}, {[Online; Accessed: 16-June-2023]}.

\bibitem{w:fitbit}
{Fitbit}, ``Fitbit {Official} {Site} for {Activity} {Trackers} \& {More},'' \url{https://www.fitbit.com/global/us/home}, {[Online; Accessed: 16-June-2023]}.

\bibitem{w:garmin}
{Garmin}, ``Running {Dynamics} {\textbar} {Garmin} {Technology},'' \url{https://www.garmin.com/en-US/garmin-technology/running-science/running-dynamics/}, {[Online; Accessed: 16-June-2023]}.

\bibitem{w:biosign}
{BioSign GmbH}, ``{BioSign} {GmbH},'' \url{https://site.biosign.de/qiu}, {[Online; Accessed: 16-June-2023]}.

\bibitem{j:janssen_2020}
T.~Janssen \emph{et~al.}, ``Lora 2.4 ghz communication link and range,'' \emph{Sensors}, vol.~20, p. 4366, 08 2020, doi: 10.3390/s20164366.

\bibitem{b:hastie_2009}
T.~Hastie, R.~Tibshirani, J.~H. Friedman, and J.~H. Friedman, \emph{The elements of statistical learning: data mining, inference, and prediction}.\hskip 1em plus 0.5em minus 0.4em\relax Springer, 2009, ISBN 978-0387848570.

\bibitem{biocas_thorir}
T.~M. Ingolfsson \emph{et~al.}, ``Towards long-term non-invasive monitoring for epilepsy via wearable eeg devices,'' in \emph{2021 IEEE Biomedical Circuits and Systems Conference (BioCAS)}, Berlin, Germany, 2021, doi: 10.1109/BioCAS49922.2021.9644949.

\bibitem{c:ingolfsson2022}
T.~M. e.~a. Ingolfsson, ``Energy-efficient tree-based eeg artifact detection,'' in \emph{2022 44th Annual Internat. Conf.e of the IEEE Eng. in Medicine \& Biology Society (EMBC)}, Glasgow, Scotland, United Kingdom, 2022, doi: 10.1109/EMBC48229.2022.9871413.

\bibitem{w:greenwaves2022}
{GreenWaves Technologies}, ``{GAP9 processor},'' \url{https://greenwaves-technologies.com/processor\_noise\_cancellation}, {[Online; Accessed: 16-June-2023]}.

\bibitem{w:mlcommons2022}
{MLCommons}, ``{MLCommons} benchmark v1.0 {Results}, {Inference}: tiny,'' \url{https://mlcommons.org/en/inference-tiny-10/}, {[Online; Accessed: 16-June-2023]}.

\bibitem{c:ingolfsson2021}
T.~M. Ingolfsson \emph{et~al.}, ``Ecg-tcn: Wearable cardiac arrhythmia detection with a temporal convolutional network,'' in \emph{2021 IEEE 3rd International Conference on Artificial Intelligence Circuits and Systems (AICAS)}, Washington DC, DC, USA, 2021, doi: 10.1109/AICAS51828.2021.9458520.

\end{thebibliography}
\end{document}